\documentclass[twocolumn,prl,aps,draft]{revtex4}
\usepackage{psfig}
\newcommand{\beq}{\begin{equation}}
\newcommand{\eeq}{\end{equation}}
\newcommand{\beqn}{\begin{eqnarray}}
\newcommand{\eeqn}{\end{eqnarray}}

\newcommand\la{\langle}
\newcommand\ra{\rangle}
\newcommand\eps\varepsilon
\def\lsim{\mathrel{\rlap{\lower4pt\hbox{\hskip1pt$\sim$}}
    \raise1pt\hbox{$<$}}}         
\def\gsim{\mathrel{\rlap{\lower4pt\hbox{\hskip1pt$\sim$}}
    \raise1pt\hbox{$>$}}}         

\begin{document}
\title{\vphantom{XX}
\Large QCD Coherence Effects in Low-x DIS and Drell-Yan\\ with Nuclear Targets}
\author{\bf J\"org Raufeisen}
\affiliation{\bf Los Alamos National Laboratory, MS H846,\\
		 Los Alamos,
		  NM 87545, USA}

\maketitle    


\section{ABSTRACT}

We investigate QCD coherence effects in deep inelastic scattering (DIS)
off nuclei
and in Drell-Yan (DY) dilepton production in proton-nucleus collisions
within the light-cone color-dipole approach. The phys\-i\-cal mechanisms 
underlying the nuclear effects become very transparent in this approach 
and are explained in some detail. We present numerical calculations of
nuclear shadowing in DIS and DY and compare to data. Nuclear effects in 
the DY transverse momentum distribution are calculated as well. The
dipole approach is the only known way to calculate the Cronin effect
without introducing additional parameters for nuclear targets.

\medskip

\noindent
{\bf Keywords:} Deep Inelastic Scattering, Drell-Yan Process, QCD, 
Nuclear Effects, Coherence Effects, Multiple Scattering.

\section{1.~INTRODUCTION}

The use of nuclei instead of protons
in high energy scattering experiments,
like deep inelastic scattering,
provides unique possibilities to
study the space-time development of strongly interacting systems. In
experiments with proton targets the products of the scattering process can only
be observed in a detector which is separated from the reaction point by a
macroscopic distance. In contrast to this, the nuclear medium can serve as a
detector located directly at the place where the microscopic interaction
happens.
As a consequence, with nuclei one can 
study coherence effects in QCD which are
not accessible in DIS off
protons nor in proton-proton scattering.

\begin{figure}[htb]
\centerline{\psfig{figure=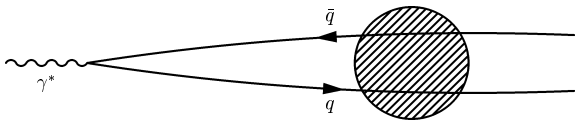,width=8cm}}
\protect\caption{At low $x_{Bj}$ and in the target rest frame, the
virtual photon $\gamma^*$ converts into a $q\bar q$-pair long before
the target.}
\label{fig:disfig}
\end{figure}

At high energies, nuclear scattering is governed by coherence effects
which are most easily understood
in the target rest frame. In the rest frame, 
DIS looks like pair creation from a virtual
photon, see Fig.\ \ref{fig:disfig}. 
Long before the target, the virtual photon splits into a $q\bar
q$-pair. The lifetime $l_c$ of the fluctuation, which is also called
coherence length,
can be estimated
with help of the uncertainty relation to be of order $\sim 1/m_N x_{Bj}$,
where $x_{Bj}$ is Bjorken-x and
$m_N\approx 1$ GeV is the nucleon mass. 
The coherence length can become
much greater than the nuclear radius at low $x_{Bj}$.
Multiple scattering within
the lifetime of the $q\bar q$ fluctuation
leads to the pronounced coherence effects observed in experiment.

The most
prominent example for a coherent interaction of more than one nucleon is the
phenomenon of nuclear shadowing, {\em i.e.} the suppression of 
the nuclear structure function $F_2^A$ with respect to the proton
structure function $F_2^p$ at low $x_{Bj}\lsim 0.1$, 
$F_2^A(x_{Bj},Q^2)/(AF_2^p(x_{Bj},Q^2))$. Here $Q^2$ is the virtuality of the 
photon.
Shadowing in low $x_{Bj}$ DIS and at high photon virtualities
is experimentally well studied by NMC \cite{NMC}.
The analogous effect for DY at low $x_2$ 
was discovered by E772 \cite{E772}.

What is the mechanism behind this suppression? 
If the coherence length is very long, as indicated in Fig.~\ref{fig:disfig},
the $q\bar q$-dipole  undergoes multiple scatterings inside the nucleus.
The physics of shadowing in DIS is most easily understood 
in a representation, in which the pair has a
definite transverse size $\rho$. As a result of color transparency
\cite{zkl,bbgg,bm}, small 
pairs interact with a small cross section, while large
pairs interact with a large cross section. 
The term "shadowing" can be taken literally in the target rest frame. Large
pairs are absorbed by the nucleons at the surface which cast a shadow on the
inner nucleons. The small pairs are not shadowed. They have equal chances
to interact with any of the nucleons in the
nucleus. From these simple arguments, one can already understand the two 
necessary conditions for shadowing. First, the hadronic fluctuation of the
virtual photon has to interact with a large cross section and second, the
coherence length has to be long enough to allow for multiple scattering.

\begin{figure}[htb]
\centerline{\psfig{figure=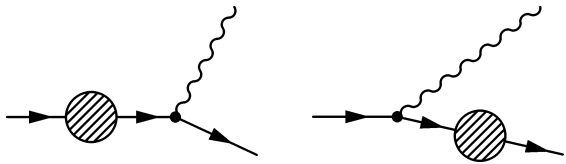,width=8cm}}
\protect\caption{A quark 
      (or an antiquark) inside the
      projectile hadron scatters off the target color field and radiates a
      massive photon. The subsequent decay of the $\gamma^*$ 
	into the lepton pair is not shown. }
\label{fig:dyfig}
\end{figure}

While the target rest frame picture of DIS is very popular, 
the light-cone approach of Kopeliovich \cite{boris,bhq,kst1}, 
which describes the DY 
process \cite{dy}
in the target rest frame, is less known.
In the light-cone approach, DY
dilepton production in the rest frame of the target appears as bremsstrahlung,
see Fig.~\ref{fig:dyfig}. A quark from the projectile scatters off the 
target and
radiates a virtual photon. This photon decays into a lepton pair. 
Even though this mechanism looks quite different from the usual annihilation
picture, it yields numerical values very similar to the next-to-leading order
parton model \cite{new}. 
Note that while cross sections are Lorentz invariant, the 
partonic interpretation of a hard scattering process is not. 
That is why a process that
looks like quark-antiquark annihilation in the infinite momentum frame can 
appear as bremsstrahlung in the target rest frame.

Remarkably,
the DY cross section can be expressed in terms of the same dipole cross section
that appears in DIS. 
In DY, the dipole cross section arises from the interference of the two
graphs in Fig.~\ref{fig:dyfig}.
On a nuclear target, the quark will of course scatter several times.
The effect of multiple scattering on
bremsstrahlung is well known in QED as the
Landau-Pomeranchuk-Migdal (LPM) effect \cite{LPM}.
The LPM effect leads to a reduction of the cross section due to destructive
interferences between the amplitudes for radiation off different points
along the quark's trajectory. In the case of DY, 
this suppression is observed in experiment as 
nuclear shadowing \cite{E772}.

\section{2.~SHADOWING IN DIS}

First, we discuss the case of shadowing in deep inelastic scattering off
nuclei.
Like shadowing in hadron-nucleus collisions \cite{glaubergribov}, 
also shadowing in DIS is intimately related to diffraction. 
The close connection between shadowing and diffraction becomes most transparent
in the formula derived by Karmanov and Kondratyuk \cite{KKK}. In the double
scattering approximation, the shadowing correction can be related to the
diffraction dissociation spectrum, integrated over the mass,
\beqn\nonumber
{\sigma^{\gamma^*A}}
&\approx& A\sigma^{\gamma^*p}
\\
\label{eq:kkk}
\nonumber
&-&{4\pi}\,
\int dM^2_X\,\left.\frac{d\sigma(\gamma^*N\to XN)}
{dM^2_X\,dt}\right|_{t\to 0}\\
&\times&\int d^2bF_A^2(l_c,b).
\eeqn
Here
\beq\label{ff}
F_A^2(l_c,b)=\left|\int\limits_{-\infty}^{\infty}dz\rho_A(b,z)
{\rm e}^{{\rm i}z/l_c}\right|^2
\eeq
is the formfactor of the nucleus, which depends on the coherence length 
\beq\label{length}
l_c=\frac{2\nu}{Q^2+M_X^2},
\eeq
$\nu$ is the energy of the $\gamma^*$ in the target rest frame
and $M_X$ is the mass of the diffractively excited state.
The coherence length can be estimated from the uncertainty relation and 
is the lifetime of the diffractively excited state. If $l_c\to 0$,
the shadowing correction in Eq.~(\ref{eq:kkk}) vanishes and one is left
with the single scattering term $A\sigma^{\gamma^*p}$. 

Note that Eq.~(\ref{eq:kkk}) is valid only in double scattering approximation.
For heavy nuclei, however, higher order scattering terms will become 
important. These can be calculated, if one knows the eigenstates of the 
interaction. Fortunately, the eigenstates of the $T$ matrix (restricted
to diffractive processes) were identified a long time ago in QCD 
\cite{zkl,mp} as partonic 
configurations with fixed transverse separations in impact parameter space.
For DIS, the lowest eigenstate is the $q\bar q$ Fock component of the
photon. The total $\gamma^*$-proton cross section is easily calculated,
if one knows the cross section $\sigma_{q\bar q}(\rho)$
for scattering a $q\bar q$-dipole of transverse size $\rho$ off a proton,
\beq\label{eq:tot}
\sigma^{\gamma^*p}=\int d\alpha d^2\rho
\left|\Psi_{q\bar q}(\alpha,\rho)\right|^2\sigma_{q\bar q}(\rho).
\eeq
The light-cone wavefunction $\Psi_{q\bar q}(\alpha,\rho)$ 
describes the splitting of the virtual photon into the $q\bar q$-pair and
is calculable 
in perturbation theory, see {\em e.g.} \cite{nz}. 
Here, $\alpha$ is the longitudinal momentum fraction carried by the quark
in Fig.~\ref{fig:disfig}.
The dipole
cross section is governed by nonperturbative effects and cannot 
be calculated from first principles. 
We use the phenomenological parameterization from 
\cite{thesis} for our calculation of shadowing in DIS, which is fitted to
HERA data on the proton structure function.
Note that higher Fock-states of the photon, containing gluons, lead
to an energy dependence of $\sigma_{q\bar q}$, which we do not write
out explicitly.

The diffractive cross section can also be expressed in terms of
$\sigma_{q\bar q}$. Since the cross section for diffraction is
proportional to the square of the $T$-matrix element, 
$|\la\gamma^*|T|X\ra|^2$, the dipole cross section also
enters squared,
\beq\label{eq:diff}
\int dM^2_X\,\left.\frac{d\sigma(\gamma^*N\to XN)}
{dM^2_X\,dt}\right|_{t\to 0}=\frac{\la\sigma^2_{q\bar q}(\rho)\ra}
{16\pi},
\eeq
where the brackets $\la\dots\ra$ indicate averaging over the light-cone
wavefunctions like in Eq.~(\ref{eq:tot}). We point out that in
order to reproduce the
correct behavior of the diffractive cross section at large $M_X$, one has to
include at least the $q\bar qG$ Fock-state of the $\gamma^*$. This correction 
is, however, of minor importance in the region where shadowing data are 
available.  

If one attempts to calculate shadowing from Eq.~(\ref{eq:kkk}) with help of
Eq.~(\ref{eq:diff}), one faces the problem that the nuclear form factor, 
Eq.~(\ref{ff}), depends on the mass $M_X$ of the diffractively produced
state, which is undefined in impact parameter representation.
Only in the limit
$l_c\gg R_A$, where $R_A$ is the nuclear radius,
it is possible to resum the entire multiple scattering series
in an eikonal-formula
\beq\label{eikonalappr}
\sigma^{\gamma^*A}=\left< 2\int d^2b\left(
1-\exp\left(-\frac{\sigma_{q\bar q}(\rho)}{2}T(b)\right)\right)\right>.
\eeq
The nuclear thickness function 
$T(b)=\int_{-\infty}^{\infty}dz\,\rho_A(b,z)$ is the
integral of
nuclear density over longitudinal
coordinate $z$ and depends on the
impact parameter $b$.
The condition $l_c\gg R_A$
insures that the $\rho$ does not vary during propagation
through the nucleus (Lorentz time dilation) and one can apply the eikonal
approximation.

\begin{figure}[htb]
\centerline{\psfig{figure=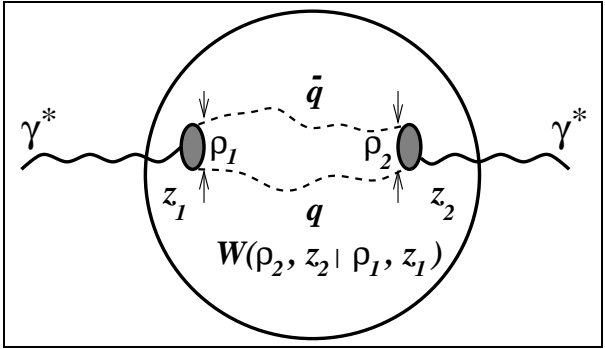,width=8cm}}
\protect\caption{Propagation of a $q\bar q$-pair through a nucleus. 
Shown is the case of
a finite coherence length, where the transverse motion is described by the
Green function $W\left(\vec\rho_2,z_2|\vec\rho_1,z_1\right)$. }
\label{fig:propag}
\end{figure}

The condition $l_c\gg R_A$ is however not fulfilled in experiment.
For the case $l_c \sim R_A$, one has to take the
variation of
$\rho$ during propagation of the $q\bar q$ fluctuation through
the nucleus into account, see Fig.~\ref{fig:propag}. A widely used recipe is
to replace $M_X^2\to Q^2$, so that $l_c\to 1/(2m_Nx_{Bj})$ and one
can apply the double scattering approximation. This recipe was, however,
disfavored during our investigation \cite{krt2}. Moreover, there
is no simple recipe to include a finite $l_c$ into higher order scattering 
terms.

In \cite{krt1} a Green function technique was developed that 
provides the correct quantum-mechanical treatment of a finite coherence length
in all multiple scattering terms.
Like in Eq.~(\ref{eq:kkk}) the total cross section 
is represented in the form
\beq\label{form}
\sigma^{\gamma^*A}=A\sigma^{\gamma^*p}-\Delta\sigma,
\eeq
where $\Delta\sigma$ is the shadowing correction,
\beqn\nonumber\label{correction}
\Delta\sigma&=&\frac{1}{2}{\rm Re}\int d^2b
\int\limits_{-\infty}^{\infty} dz_1\rho_A(b,z_1)
\int\limits_{z_1}^{\infty} dz_2\rho_A(b,z_2)\\
&\times&\nonumber
\int_0^1d\alpha\int d^2\rho_2
\Psi_{q\bar q}^*(\vec\rho_2,\alpha)\sigma_{q\bar q}(\rho_2)\\
&\times&
A(\vec\rho_2,z_1,z_2,\alpha),
\eeqn
with
\beqn\label{propagation}
A(\vec\rho_2,z_1,z_2,\alpha)&=&\int\,d^2\rho_1\,
W(\vec \rho_2,z_2|\vec \rho_1,z_1)\,\\
\nonumber
&\times&
{\rm e}^{-{\rm i}q_L^{min}(z_2-z_1)}\,\sigma_{q\bar q}(\rho_1)\,
\Psi_{q\bar q}(\vec \rho_1,\alpha).
\eeqn
Here,
\beq
q_L^{min}=\frac{1}{l_c^{max}}=\frac{Q^2\alpha(1-\alpha)
+m_f^2}{2\nu\alpha(1-\alpha)}
\eeq
is the minimal longitudinal momentum transfer when the photon splits into the
$q\bar q$ dipole ($m_f$ is the quark mass). 

The shadowing term in Eq.~(\ref{form}) is
illustrated
in Fig.~\ref{fig:propag}.  
At the point $z_1$ the photon diffractively produces the $q\bar q$
pair ($\gamma^*N\to q\bar qN$) with a transverse separation $\vec\rho_1$.
The pair propagates through the nucleus along arbitrarily curved
trajectories, which are summed over, and arrives at the point
$z_2$ with a separation $\vec\rho_2$.  The initial and the
final separations are controlled by the light-cone wavefunction 
$\Psi_{q\bar q}(\vec \rho,\alpha)$.  While passing the nucleus
the $q\bar q$ pair interacts with bound nucleons
via the cross section $\sigma_{q\bar q}(\rho)$ which depends on the local
separation $\vec\rho$.  The Green function 
$W(\vec\rho_2,z_2|\vec\rho_1,z_1)$ describes the propagation of the pair from
$z_1$ to $z_2$, see Eq.~(\ref{propagation}), including all
multiple rescatterings and a finite coherence length.
Note the diffraction dissociation ($DD$) amplitude,
\beq
f_{DD}(\gamma^*\to q\bar q)={\rm i}\Psi_{q\bar q}(\vec \rho_1,\alpha)
\sigma_{q\bar q}(\rho_1)
\eeq,
in Eq.~(\ref{propagation}).
At the position $z_2$, the result of the propagation is again 
projected onto the
diffraction dissociation amplitude, Eq.~(\ref{correction}). 
The Green function
includes that part of the phase shift between the
initial and the final photons which is due to transverse
motion of the quarks, while the longitudinal motion is included in
Eq.~(\ref{propagation}) via the exponential.

The Green function $W(\vec \rho_2,z_2;\vec\rho_1,z_1)$ in
Eq.~(\ref{propagation}) satisfies the two dimensional 
Schr\"o\-din\-ger equation,
\beqn\label{schroedinger}\nonumber\lefteqn{
{\rm i}\,\frac{{\partial}W(\vec\rho_2,z_2|\vec\rho_1,z_1)}{{\partial}z_2}
=}\\
&-&\frac{\Delta(\rho_2)}{2\nu\alpha(1-\alpha)}\,
W(\vec\rho_2,z_2|\vec\rho_1,z_1)\nonumber\\
&-&{{\rm i}\over2}\,\sigma(\rho_2)\,\rho_A(b,z_2)\,
 W(\vec\rho_2,z_2|\vec\rho_1,z_1)
\eeqn
with the boundary condition $W(\vec\rho_2,z_1|\vec\rho_1,z_1)
=\delta^{(2)}(\vec \rho_2-\vec \rho_1)$.  The Laplacian $\Delta(\rho_2)$ acts
on the coordinate $\vec\rho_2$.   
The kinetic term $\Delta/[2\nu\alpha(1-\alpha)]$ in this Schr\"odinger equation
takes care of the varying effective mass of the $q\bar q$ pair
and provides the proper phase shift.
The role of time is played by the
longitudinal coordinate $z_2$.
The imaginary part of the optical potential describes the rescattering.

The Green function method contains the eikonal
approximation Eq.~(\ref{eikonalappr}) and the Karmanov-Kondratyuk formula 
Eq.~(\ref{eq:kkk}) as limiting cases.
In order to obtain the eikonal approximation, one has to take the limit
$\nu \to\infty$. In this case, the kinetic energy term in
Eq.~(\ref{schroedinger}) can be neglected and  
with $q_L^{min}\to 0$ one arrives 
after a short calculation at  Eq.~(\ref{eikonalappr}).
 
One can also recover the Karmanov-Kondratyuk formula, 
when one neglects 
the imaginary potential in Eq.~(\ref{schroedinger}). 
Then $W$ becomes the  Green function of a free motion.

\begin{figure}[htb]
\centerline{\psfig{figure=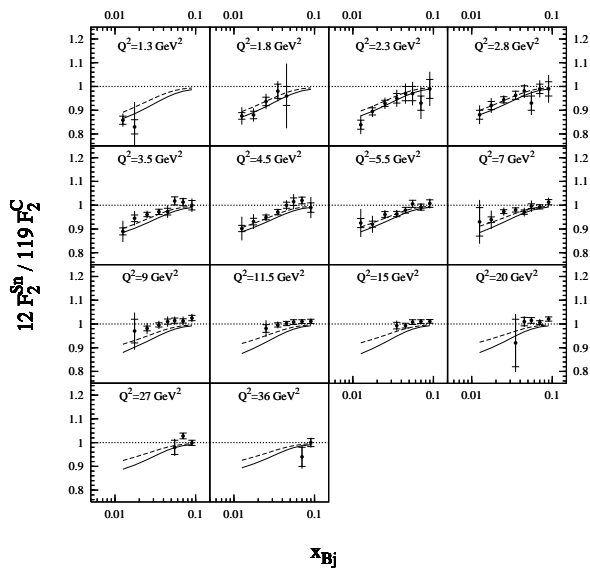,width=8cm}}
\protect\caption{The $x_{Bj}$ dependence of shadowing in DIS
for the structure function of tin relative to carbon. The data are 
from NMC \cite{NMC}. The full curves are calculated in the Green-function
approach, including the nonperturbative interaction between the $q$ and the 
$\bar q$. The dashed curve does not include this interaction. The figure is 
from \cite{thesis}. }
\label{fig:nmc}
\end{figure}

Calculations in the Green function approach are compared to NMC data in
Fig.~\ref{fig:nmc}. Note that the leading twist contribution to shadowing 
is due to large dipole sizes, where nonperturbative effects, like 
an interaction between the $q$ and the $\bar q$, might become important.
Therefore, the solid curve is calculated following \cite{kst2}, including such
an interaction. This interaction modifies the light-cone wavefunction 
\cite{kst2}.
The dashed curve is calculated with the conventional, perturbative light-cone
wavefunctions, but including a constituent quark mass. Both curves are 
in reasonable agreement with the data. 

Note that 
for these data, the coherence length is of order of the
nuclear radius or smaller. Indeed, shadowing vanishes around 
$x_{Bj}\approx 0.1$, because the coherence length becomes smaller than
the mean internucleon spacing. Therefore, the eikonal approximation, 
Eq.~(\ref{eikonalappr}), cannot be applied for the kinematics of NMC
and a correct treatment of the coherence length becomes crucial.
We emphasize that the calculation in Fig.~\ref{fig:nmc} does not 
contain any free parameters. Following the spirit of Glauber theory,
all free parameters are adjusted to DIS off protons.

\section{3.~NUCLEAR EFFECTS IN THE DRELL-YAN PROCESS}

In the second part of this paper, we discuss the application of the
dipole approach to calculate nuclear effects in DY. The main 
features of the approach have already been introduced in the preceding 
section and the case of DY is almost completely analogous to DIS.

The cross section for radiation of a virtual photon from a quark after
scattering on a proton, can be written in factorized light-cone form
\cite{boris,bhq,kst1}, 
\beq\label{dylctotal}
\frac{d\sigma(qp\to \gamma^*X)}{d\ln\alpha}
=\int d^2\rho\, |\Psi_{\gamma^* q}(\alpha,\rho)|^2
    \sigma_{q\bar q}(\alpha\rho),
\eeq
with the same $\sigma_{q\bar q}$ as in DIS. Again, the light-cone
wavefunction $\Psi_{\gamma^* q}(\alpha,\rho)$ is calculable 
in perturbation theory \cite{krt3,new}.
For DY, however,
$\rho$ is  the photon-quark transverse separation and $\alpha$ 
is the fraction of 
the light-cone momentum of the initial quark taken away by the photon.
We use the standard notation for the kinematical variables,
$x_1-x_2=x_F$, $\tau=M^2/s=x_1x_2$, where $x_F$ is the
Feynman variable,
$s$ is the center of mass energy squared of the colliding protons and 
$M$ is the
dilepton mass. 

The physical interpretation of (\ref{dylctotal}) is similar to the DIS
case. Long before the target, the projectile quark develops fluctuations 
which contain virtual photons. On interaction with the target, these 
fluctuations can be put on mass shell and the $\gamma^*$ is freed.
In order to make this happen, the interaction must distinguish 
between Fock states with and without $\gamma^*$, {\it i.e.} 
these states have to interact differently.
Since only the quark
interacts in both Fock components, the difference in the scattering
amplitudes arises from the  
relative displacement of the quark
in the transverse plane when it radiates a photon.
The $\gamma^*q$ fluctuation has a center of 
gravity in the
transverse plane which coincides with the impact parameter of the parent quark.
One finds that the transverse separation between the quark and the center of 
gravity is
$\alpha\rho$, {\em i.e.} the argument of $\sigma_{q\bar q}$.
The dipole cross section arises from the interference of the two graphs in
Fig.~\ref{fig:dyfig} and should be interpreted as freeing cross section
for the virtual photon.
More discussion can be found in \cite{krt3}.

The transverse momentum distribution of DY pairs
can also be expressed in terms of the dipole cross section \cite{kst1}. 
The differential cross section is given by the 
Fourier integral
\beqn\nonumber\label{dylcdiff}\lefteqn{
\frac{d\sigma(qp\to \gamma^*X)}{d\ln\alpha d^2q_{T}}=}\\
\nonumber
&=&\frac{1}{(2\pi)^2}
\int d^2\rho_1d^2\rho_2\, \exp[{\rm i}\vec q_{T}\cdot(\vec\rho_1-\vec\rho_2)]
\\
&\times&\Psi^*_{\gamma^* q}(\alpha,\vec\rho_1)
\Psi_{\gamma^* q}(\alpha,\vec\rho_2)\\
\nonumber
&\times&
\frac{1}{2}
\left\{\sigma_{q\bar q}(\alpha\rho_1)
+\sigma_{q\bar q}(\alpha\rho_2)
-\sigma_{q\bar q}(\alpha(\vec\rho_1-\vec\rho_2))\right\}.
\eeqn
After integrating this expression over the transverse momentum
$q_{T}$ of the photon, one obviously recovers
Eq.~(\ref{dylctotal}). 

Nuclear effects arise in the DY process, because the different nucleons in the
nucleus compete to free the $\gamma^*$. Necessary condition for this
phenomenon is of course that the coherence length for DY,
\beq
l_c=\frac{1}{m_Nx_2}\frac{(1-\alpha)M^2}{q_T^2+(1-\alpha)M^2+\alpha^2m_f^2}
\eeq
exceeds the mean internucleon spacing of $\sim 2$ fm.
At infinite coherence length, one can simply eikonalize the 
dipole cross section, similar to the DIS case, cmp.\ 
Eq.~(\ref{eikonalappr}).
For finite coherence length, the Green function technique
can be developed for DY in complete analogy to DIS. We do not display the
corresponding formula here, which can be found in \cite{kst1,thesis,moriond}.
We point out that even though DIS and DY appear very similar in the 
dipole approach, there are no (or only very few) diffractively produced DY
pairs, because Eq.~(\ref{eq:diff}) is not applicable to an exclusive channel
(see discussion in \cite{kst2}).

\begin{figure}[htb]
\centerline{\psfig{figure=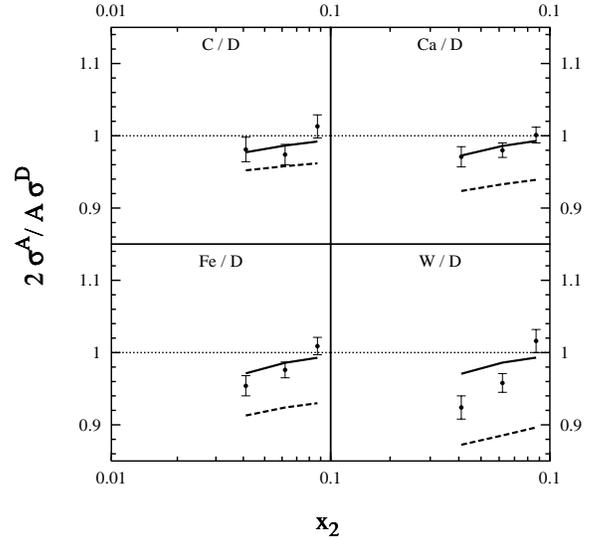,width=8cm}}
\protect\caption{\label{fig:e772}
      Comparison between calculations in the Green function 
technique (solid curve) and E772 data at 
center of mass energy {$\sqrt{s}=38.8$ GeV} 
  for shadowing in DY. The dashed curve shows the eikonal 
approximation, 
which is
not valid at this energy, any more. }
\end{figure}

Calculations of shadowing for DY
in the Green function approach are compared to E772 data 
\cite{E772} in Fig.~\ref{fig:e772}. Like in the case of the NMC data
in the previous section,
the coherence length $l_c$ at E772 
energy becomes smaller than the nuclear radius. Shadowing vanishes as 
$x_2$ approaches $0.1$, because the coherence length becomes smaller 
than the mean internucleon separation. Again,
it is therefore important 
to have a correct description of a finite $l_c$. 
The eikonal approximation does not reproduce 
the vanishing shadowing toward $x_2\to 0.1$, because it assumes an infinite
coherence length.
 
The curves in  Fig.~\ref{fig:e772} are somewhat different from the ones in
\cite{thesis,moriond}, because we used a different parameterization of the
dipole cross section \cite{wuesthoff}.
Note that for heavy nuclei, energy loss \cite{eloss}
leads to an additional
suppression of the DY cross section. It is therefore important to
calculate shadowing in DY without introducing additional parameter for nuclear
targets. Otherwise, shadowing and energy loss cannot be disentangled.

\begin{figure}[b]
\centerline{\psfig{figure=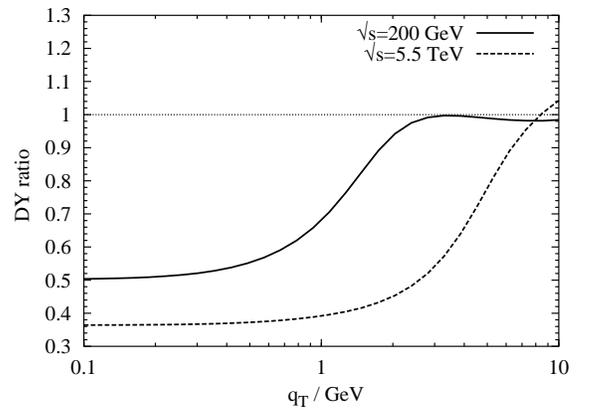,width=8cm}}
\protect\caption{\label{fig:qt_ratio}
  Nuclear effects on the DY transverse momentum distribution at 
  RHIC and LHC for
  dilepton mass $M=4.5$ GeV and Feynman $x_F=0.5$.  }
\end{figure}

Nuclear effects on the
$q_{T}$-differential cross section calculated at RHIC and LHC energy
are shown in Fig.~\ref{fig:qt_ratio}.
At these high {\em cm} energies, the coherence length 
significantly exceeds the nuclear radius, justifying the 
eikonal approximation. However,
in the case of such long $l_c$, the quark has sufficient
time to radiate additional gluons. Thus, higher Fock states have
to be taken into account. This is accomplished by multiplying 
$\sigma_{q\bar q}$ with the gluon shadowing ratio $R_G$, which was calculated
in \cite{kst2}. See \cite{krtj} for a detailed discussion.

The differential cross section is suppressed at small
transverse momentum $q_{T}$ of the dilepton, 
where large values of $\rho$ dominate. This suppression
vanishes at intermediate $q_{T}\sim 2$ GeV. 
The Cronin enhancement that one could expect in this intermediate
$q_T$ region \cite{moriond}
is suppressed due to gluon shadowing \cite{krtj}.  
At very large 
transverse momentum nuclear
effects vanish.

\section{4.~SUMMARY}

In the target rest frame and at low $x_{Bj}$ ($x_2$),
the cross sections for DIS and DY can both be expressed in terms
of the same cross section for scattering a color-neutral $q\bar q$-pair
off a proton. The advantage of this formulation and the main motivation
to develop this approach is the easy generalization to nuclear targets
and the insight into the dynamical origin of nuclear effects, which 
appear as coherence effects due to multiple scattering.

The main nonperturbative input to all formulae is the dipole cross section,
which cannot be calculated from first principles. Instead one uses a
phenomenological parameterization for this quantity, which is determined
to fit DIS data from HERA. Nuclear effects are then calculated without 
introducing any new parameters. A parameter free calculation of nuclear 
shadowing for DY is indispensable, if one aims at extracting the energy loss
of a fast quark propagating through nuclear from DY data \cite{eloss}.
Furthermore, the dipole formulation is the only known way to calculate the
Cronin effect in a parameter free way. This approach can also be applied to
calculate the Cronin effect in hadronic collisions \cite{ce}.
Furthermore, a good description of NMC
data on shadowing in DIS is achieved. 

Finally, we mention that 
the dipole approach can also be applied to a variety of other processes,
{\em e.g.} heavy quark, quarkonium or vector meson production \cite{bk}.

\vspace{2mm}

\noindent {\bf Acknowledgments}: This work was supported by the
U.S.~Department of Energy at Los Alamos National Laboratory under Contract
No.~W-7405-ENG-38. The author is grateful to J\"org H\"ufner, Mikkel Johnson,
Boris Kopeliovich, Jen-Chieh Peng 
and Alexander Tarasov for valuable discussion.

\end{document}